\documentclass[useAMS,usenatbib, preprint]{mn2e}
\usepackage[latin1]{inputenc}
\usepackage{txfonts}
\usepackage[dvips]{graphicx}
\usepackage{natbib}
\bibpunct{(}{)}{;}{a}{}{,}
\usepackage{float}
\usepackage{color}
\usepackage{cancel}
\usepackage{graphicx}
\usepackage{xcolor}
\usepackage{ulem}

\newcommand{\be}{\begin{equation}}
\newcommand{\ee}{\end{equation}}
\newcommand{\bea}{\begin{eqnarray}}
\newcommand{\eea}{\end{eqnarray}}
\newcommand{\ba}{\begin{array}}
\newcommand{\ea}{\end{array}}

\newcommand{\bit}{\begin{itemize}}
\newcommand{\eit}{\end{itemize}}
\newcommand{\ben}{\begin{enumerate}}
\newcommand{\een}{\end{enumerate}}

\newcommand{\aap}{    {\it Astron. Astrophys.}}

\newcommand{\apj}{    {\it Astrophys. J.}}
\newcommand{\apjl}{   {\it Astrophys. J. Lett.}}

\newcommand{\grl}{    {\it Geophys. Res. Lett.}}

\newcommand{\jastp}{  {\it J. Atmos. Sol. Terr. Phys.}}

\newcommand{\mnras}{  {\it Mon. Not. Roy. Astron. Soc.}}
\newcommand{\nat}{    {\it Nature}}

\newcommand{\prl}{    {\it Phys. Rev. Lett.}}
\newcommand{\solphys}{{\it Solar Phys.}}

\begin{document}

\title{Effects of cyclic fluctuations in meridional circulation using a low order dynamo model}

\author[Passos \& Lopes]{D. Passos$^{1,2,3}$, I. Lopes$^{1,3}$\\
$^{1}$ Departamento de F\'\i sica,
Universidade de \'Evora, Col\'egio Ant\'onio Luis
Verney, 7002-554 \'Evora - Portugal,\\
$^{2}$ GRPS, D\'epartment de Physique, Universit\'e de Montr\'eal, C.P. 6128
Centre-ville, Montr\'eal, Qc, Canada  H3C-3J7 \\
$^{3}$ CENTRA-IST, Instituto Superior
T\'ecnico, Av. Rovisco Pais, 1049-001 Lisboa, Portugal,\\
}

\maketitle

\label{firstpage}

\begin{abstract}

We develop and subsequently explore the solution space of a simple flux transport dynamo model that incorporates a time dependent large scale meridional circulation. Based on recent observations we prescribed an analytical form for the amplitude of this circulation and study its impact in the evolution of the magnetic field. We find that cyclic variations in the amplitude and frequency of the meridional flow affect the strength of the solar cycle. Variations in the amplitude of the fluctuations influence the shape of the solar cycle but are only relevant to the cycle's strength variations when they occur at a frequency different from or out of phase of the solar cycle's.
\end{abstract}

\begin{keywords}
Magnetic fields; Solar activity cycle; Dynamo models; Meridional circulation
\end{keywords}

%%%------------------------------------ INTRO ----------------------------------

\section{Introduction}

Our civilization is increasingly becoming more and more dependent on energy distribution, communication networks and satellite operation since these technologies provide an important backbone for our daily activities. Nevertheless, these key technological assets may face dangers that one would like to minimize. Our Sun, so important for the life on our planet, is now being pointed as a source of problems for these technologies. Solar magnetic storms can surreptitiously hit Earth and damage all the structures previously mentioned. At the origin of these storms we can find the large scale solar magnetic field. This field is believed to be originated by a magnetohydrodynamic dynamo that converts kinetic energy from the solar plasma motions into magnetic energy~\citep{Parker1955}. Several dynamo models have been developed to explain the main observational features of the solar magnetic field~\citep{Jouve2008, Charbonneau2010}. These models can reproduce the magnetic field polarity reversals (every 11 years) and many field spatial features. Touted as being the most promising of the several existing types, flux transport dynamo models have been tentatively used for the first time as a tool for predictions of the solar cycle~\citep{Dikpati2006, Choudhuri2007}. These predictions were the firsts to use full dynamo models to forecast the behavior of the solar cycle. The models solve the mean-field axisymmetric equations for the magnetic field evolution on a background structure that incorporates parameterized physical mechanisms such as solar rotation, magnetic diffusivity, etc. The denomination "flux transport models" comes from the fact that they incorporate the solar meridional circulation, a conveyor belt-like plasma flow that carries magnetic field from the equator to the poles near the surface and from the poles towards the equator in the base of the solar convection zone~\citep{Dikpati1999, Chatterjee2004, Munoz-Jaramillo2009}. The amplitude or strength of this circulation controls the period and amplitude of the produced magnetic field~\citep{Nandy2002, LopesPassos2009a}. With just a few exceptions~\citep{Rempel2006, KarakChou2011}, most of these models work on the kinematic regime, which implies that the amplitude of the meridional circulation is not affected by the electromagnetic Lorentz force feedback of the magnetic field on the flow. Observational evidence of meridional circulation are very hard to obtain and current values for the surface poleward average velocity is centered around 10 to 20 ms$^{-1}$. The most reliable data is available only for the last couple of decades and the recent measurements of~\citet{Hathaway2010a} indicate that the strength of this flow might have changed by about 25\% from the beginning of cycle 23 to cycle 24. Experimental evidences for a variable meridional circulation were already reported by~\citep{Komm1993}. Also, a recent inversion methodology based in a simplified dynamo model and the annual sunspot time series proposed by~\citet{PassosLopes2008, Passos2012} suggests that the amplitude of the meridional circulation changes significantly from cycle to cycle and that those variations could explain (partially) the observed solar variability. This information is not usually taken into account in dynamo based predictions but its relevance is now recognized and is starting to be addressed by some research  groups~\citep{Karak2010, Hotta2010, Nandy2011}. It has been shown by~\citet{Mininni2000, WilmotSmith2005} and~\citet{PassosLopes2008, PassosLopes2011} that a truncated version of the flux transport dynamo equations set or low order dynamo models can be used as a first order approximation to study the temporal behavior of the solar magnetic field. These 1D truncated models allow to calculate the evolution of the magnetic field strength by taking into account the main physical mechanisms in a simplified way in the kinematic regime, without the need to incur in heavy numerical calculations. Exploratory studies made with these models are very useful for identifying and to help focusing in the relevant aspects and physical mechanisms that should be studied in depth by 2D numerical models.  In this work we build upon the low order dynamo presented in~\citet{PassosLopes2011} by taking into account a time dependent meridional circulation's amplitude profile. After the derivation of the new low order model equations, we parameterize the meridional circulation based on the observations of~\citet{Hathaway2010b} and study the impact of this time dependence in this dynamical system's solution. The final section is dedicated to comments and remarks about the results.

\section{The model}

As previously mentioned, the model presented here is a variation of the kinematic low order dynamo model developed in~\citet{PassosLopes2008} and~\citet{PassosLopes2011}. In this version we consider that the amplitude of the meridional circulation is time dependent and the derivation of the low order model follows this directive. We start with the equations for a flux transport mean field axisymmetric dynamo as shown in~\citet{Charbonneau2010}. These equations give us the evolution of the mean solar magnetic field,  $\bar{\textbf{B}}=\textbf{B}_\phi + \textbf{B}_p$, classically decomposed into its toroidal, $\textbf{B}_\phi$, and poloidal, $\textbf{B}_p = \nabla \times (A_p \hat{e}_\phi$) components with $A_p$ representing a potential vector field.

\bea
    \frac{\partial B_\phi}{\partial t} &=& - \bar{r} \, \textbf{v}_p(t) \cdot \nabla
    \left(\frac{B_\phi} {\bar{r}}\right) + \bar{r}\left[\nabla \times (A_p \hat{e}_\phi)
    \right] \cdot \nabla \Omega\ \nonumber \\
    && + \eta \left(\nabla^2 - \frac{1}{\bar{r}^2}\right) B_\phi
    - \Gamma(B_\phi) B_\phi\,\,\, ,
    \label{eq-3} \\
    \frac{\partial A_p}{\partial t} &=& -\frac{1}{\bar{r}} \,\textbf{v}_p(t) \cdot
    \nabla \left(\bar{r}\,A_p\right) + \alpha B_\phi \nonumber \\
    && + \eta \left( \nabla^2
    - \frac{1}{\bar{r}^2} \right) A_p \,\,\, ,
    \label{eq-4}
\eea
where we have $\bar{r}=r \sin \theta$, $\nabla \Omega$ represents the differential rotation of the Sun, $\textbf{v}_p$ is the flow in the meridional plane and $\eta$ is the magnetic turbulent diffusivity. One of the simplifications used in this model is to assume an average magnetic diffusivity for the entire convection zone ($\partial \eta / \partial r = 0$) and plasma incompressibility. Following the suggestions found in~\citet{Kitchatinov2000} and~\citet{Pontieri2003} we add an extra term, $\Gamma \sim \gamma B_\phi^2 / 8 \pi \rho$, to account for magnetic flux removal by magnetic buoyancy. Here $\gamma$ is a constant related to the buoyancy regime and $\rho$ is the plasma density. As usual in these models the regeneration mechanism from toroidal to poloidal field, the so called $\alpha$-effect is represented by $\alpha$. In this case, for simplicity, we do not consider any non-linearity in $\alpha$. In the following steps we assume that $\textbf{v}_p$ depends explicitly on time and we use the dimensional approach suggested by~\citet{Mininni2000} to truncate the dynamo equations by substituting $\nabla \rightarrow 1/\ell_0$, where $\ell_0$ is a specific length of interaction for the magnetic fields, usually taken as $\ell_0\sim0.1R_\odot$. This truncation ensures that we are bounding our solution space to magnetic phenomena that occur in the scale of $\ell_0$, presumably the large scale solar magnetic field responsible for the solar cycle. After grouping terms in $B_\phi$ and $A_p$ we get
\bea
    \frac{d B_\phi}{d t} &=& \left[c_{1} -\frac{v_p(t)}{\ell_0}\right] B_\phi + c_{2} A_p - c_{3} B_\phi^3
    \label{eq-5} \\
    \frac{d A_p}{d t} &=& \left[c_{1} -\frac{v_p(t)}{\ell_0}\right] A_p + \alpha  B_\phi
    \label{eq-6}
\eea
where we have defined the coefficients, $c_n$, as
\bea
    c_{1}&=&\eta \left(\frac{1}{\ell_0^2} -
    \frac{1}{R_\odot^2} \right)
    \label{eq-7} \\
    c_{2}&=& \frac{\bar{r} \Omega}{\ell_0^2}
    \label{eq-8} \\
    c_{3} &=& \frac{\gamma}{8 \pi \rho}
    \label{eq-9}
\eea
These coefficients now contain all the structure parameters in the model, i.e. $c_1$, $c_2$ and $c_3$ assume the role of magnetic diffusivity, rotation and buoyancy respectively in this low order dynamo model. Next, we take the derivative of equation (\ref{eq-5}) with respect to time and drop the $A_p$ dependence by substituting equation (\ref{eq-6}) in the terms with $\frac{d A_p}{d t}$ and by noting that $c_2 A_p$ can be extracted from equation (\ref{eq-5}). After this mathematical workout we finally get that
\bea
    \frac{d^2 B_\phi }{d t^2} &=&
    \left[ 2 \left( c_1 - \frac{v_p(t)}{\ell_0}\right) - 3 c_3 B_\phi^2 \right] \frac{d B_\phi}{d t} \nonumber \\
    &-& \left[ \frac{1}{\ell_0} \frac{d v_p(t)}{d t} + \left(c_1 - \frac{v_p(t)}{\ell_0}\right)^2 - c_2 \alpha \right] B_\phi \nonumber \\
    &+& c_3 \left( c_1 - \frac{v_p(t)}{\ell_o}\right) B_\phi^3\,\,,
    \label{eq-10}
\eea
with $\alpha\neq0$.
The solution's space of this dynamical system is defined by the structural coefficients $c_n$ and by the analytical form of $v_p(t)$. As a note, it is important to refer that in this kind of reduced systems, the units in which some of the quantities are presented do not always coincide with the real units, rendering the direct application of known physical values troublesome. Nevertheless studies based on the relative variation of these parameters can be done and this work follows that line of thought.

\begin{figure}%[htb!]
    \centering
    \includegraphics[width=85mm]{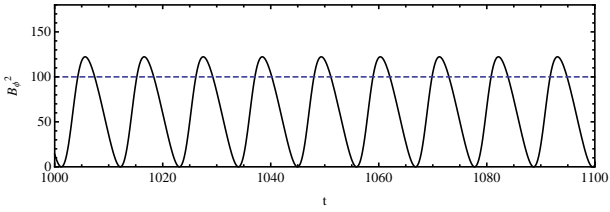}
    \caption{Static solution (obtained for constant $v_p$), after the system evolved for some time into stability. The solid black line represents $B_\phi^2$, our chosen proxy to represent the solar cycle and the dashed blue line is the scaled amplitude of $v_p/\ell_0$ (blue) in order to plot both quantities in the same scale (in this example the scaling used is 1000 $|v_p|$). Values used here $c_1=-0.01$, $c_2\alpha=-0.095$, $c_3=0.002$, $v_p/\ell_0$=-0.1.}
    \label{fig-1}
\end{figure}

A "static" $v_p$ reference solution for equation (\ref{eq-10}) is calculated by assuming a constant meridional flow amplitude. The values used for the coefficients $c_n$ in this static solution were found by fitting equation (\ref{eq-10}) to the constructed proxy for the toroidal field presented in \citet{PassosLopes2008} (for further details about the methodology used you can also see \citet{LopesPassos2009a}). Using these $c_n$ values the system presents a solar-like solution with $B_\phi^2$, a proxy representation of the solar cycle, showing a cyclic behavior with a period of about 11 years (figure (\ref{fig-1})). In the parameter regime used in this work, $v_p(t)$ behaves mathematically as a source term and $c_1$, the diffusivity as a sink term. The former is one order of magnitude higher than the latter. Also, in this parameter regime, magnetic flux removal by buoyancy, $c_3$, is the main saturation mechanism in place to avoid field growth. In the following of this work we bound ourselves to the study of variations in $v_p(t)$ maintaining the other coefficients with the values presented in the static solution. A complete study of the full parameters space is therefor deferred to a future work but some preliminary results are presented in the conclusions section.

\smallskip

\subsection{Introducing cyclic fluctuations in $v_p(t)$}

\begin{figure}%[htb!]
    \centering
    \includegraphics[width=85mm]{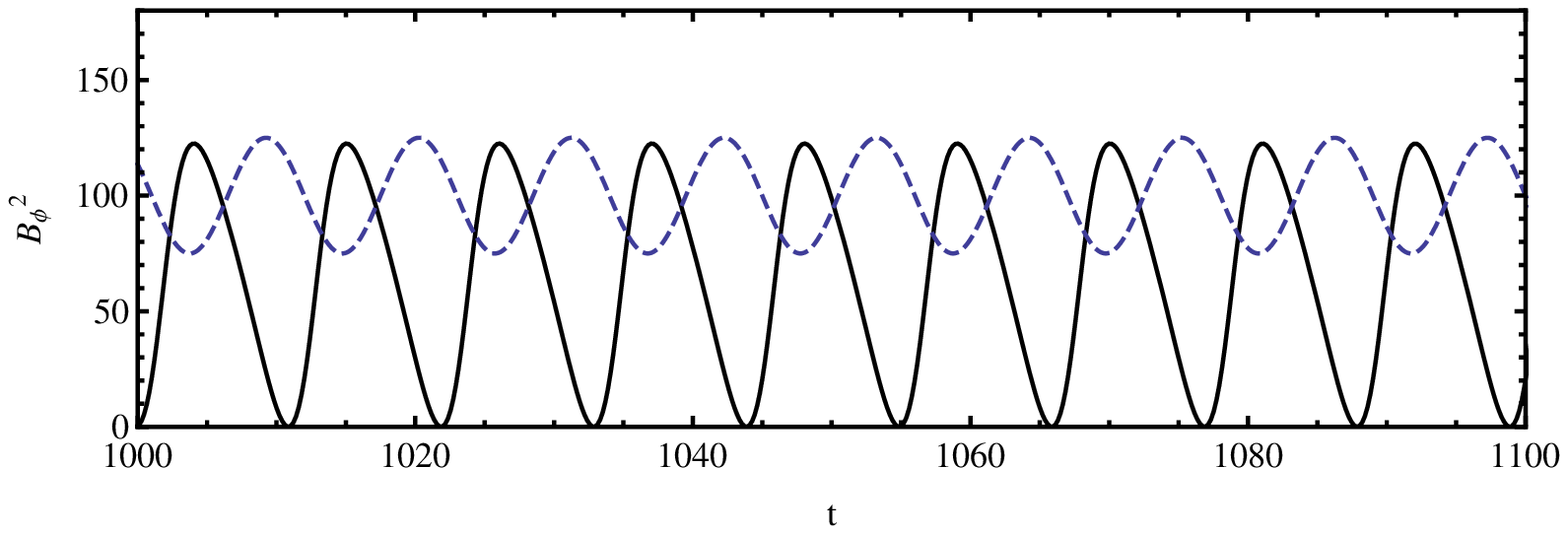}\\
    \includegraphics[width=85mm]{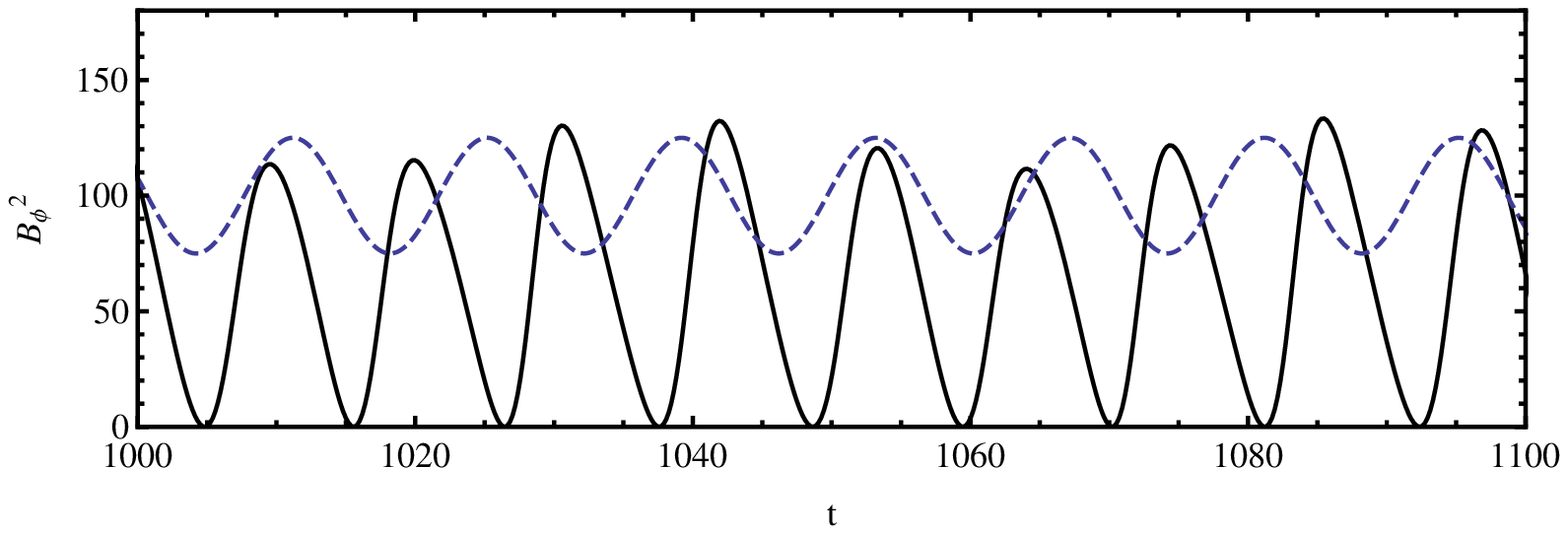}\\
    \includegraphics[width=40mm]{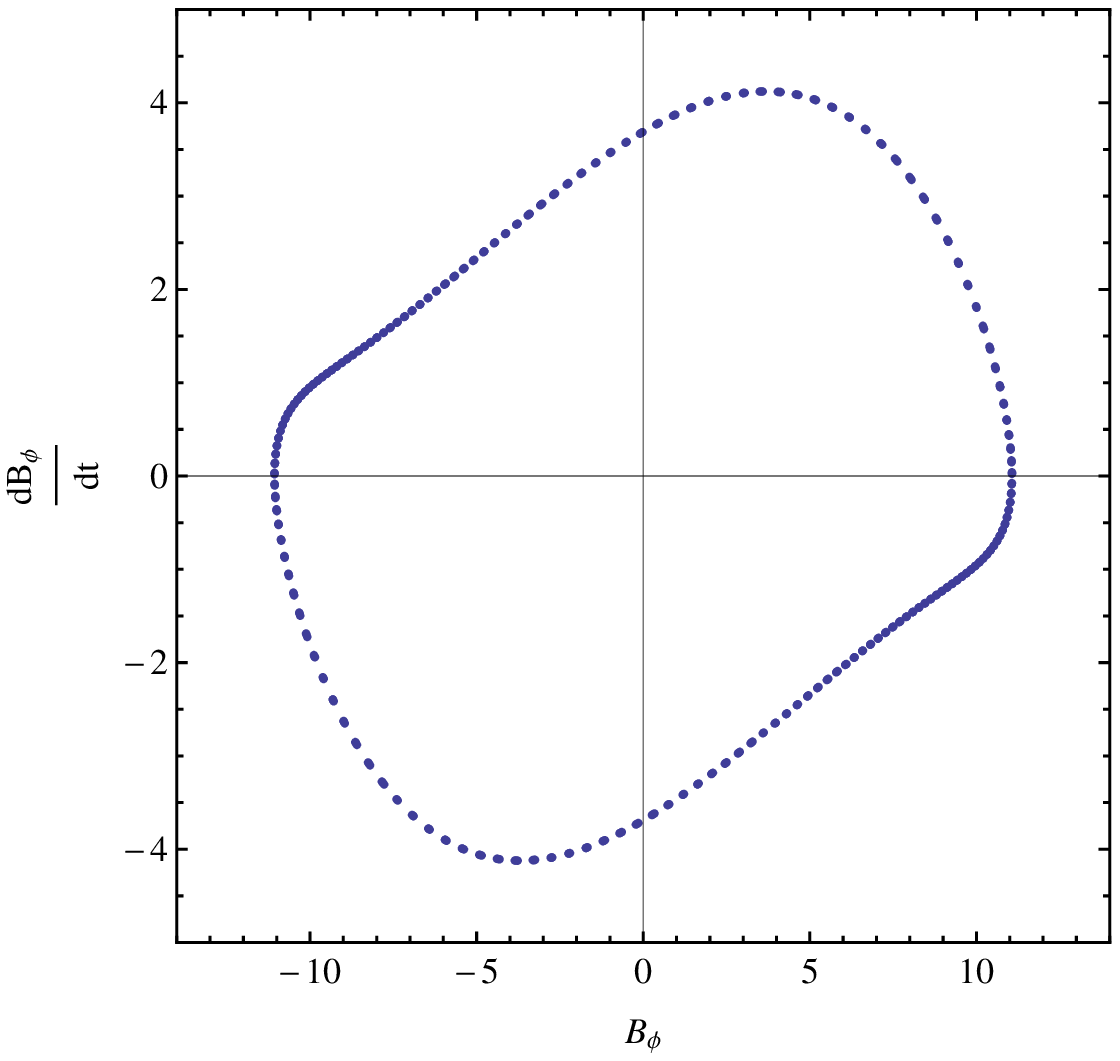}
    \includegraphics[width=40mm]{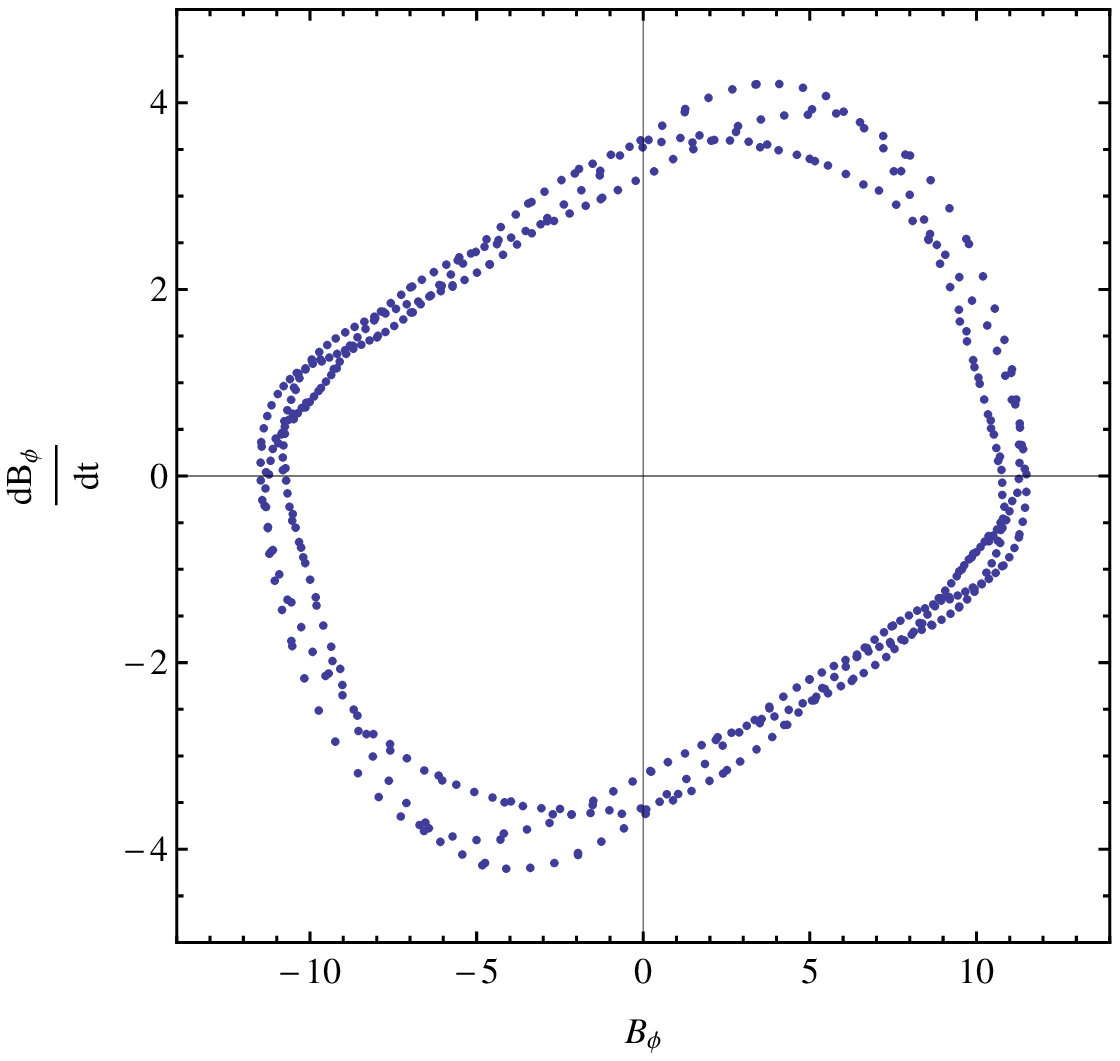}
    \caption{The lines representation is the same as in figure (1).
    Top: $c_1=-0.01$, $c_2 \alpha=-0.095$, $c_3=0.002$, $v_{p0}$=-0.1, A=0.0225, $\omega=2\pi/11$ and $\theta=0$.
    Middle: Same parameters as top panel, but with a change of the fluctuation's period, $\omega=2\pi/14$.
    In the bottom panel are represented the phase space {$B_\phi$, d$B_\phi$/d$t$} for both solution with T=11 (left) and T=14 (right) sampled at regular intervals.}
    \label{fig-2}
\end{figure}

According to the latest measurements of the surface meridional flow amplitude spanning a full magnetic cycle from \citet{Hathaway2010a}, $v_p(t)$ varies in a roughly sinusoidal way, reaching a maximum amplitude near the half of the decreasing part of the solar cycle and dropping to its lower value near the sunspot maximum (see figure (4) of \citet{Hathaway2010a}). Based in this result we propose an Ansatz where we take $\frac{v_p(t)}{\ell_0} \propto v_{p0} +$ A $\sin(\omega t + \theta)$ where A is the amplitude of the meridional flow fluctuations, $\omega$ is the frequency of these fluctuations and $\theta$ is used to control the initial phase. This parameterization translates in a fluctuation of $v_p$ around a mean value of $v_{p0}$ (value used in the static solution). The top panel of figure (\ref{fig-2}) shows the solution for $\omega = \frac{2\pi}{T}$ with $T=11$ years and a fluctuation's amplitude A=0.025 which corresponds to an amplitude variation of 25\% $v_{p0}$ (roughly the amplitude variation between cycle minimum and maximum for solar cycle 23 presented in \citet{Hathaway2010a}). It is important to mention that even if our meridional circulation is time dependent, the model is still operating in the kinematic regime since $v_p$ does not depend on the magnetic field.

\begin{figure}%[htb!]
    \centering
    \includegraphics[width=40mm]{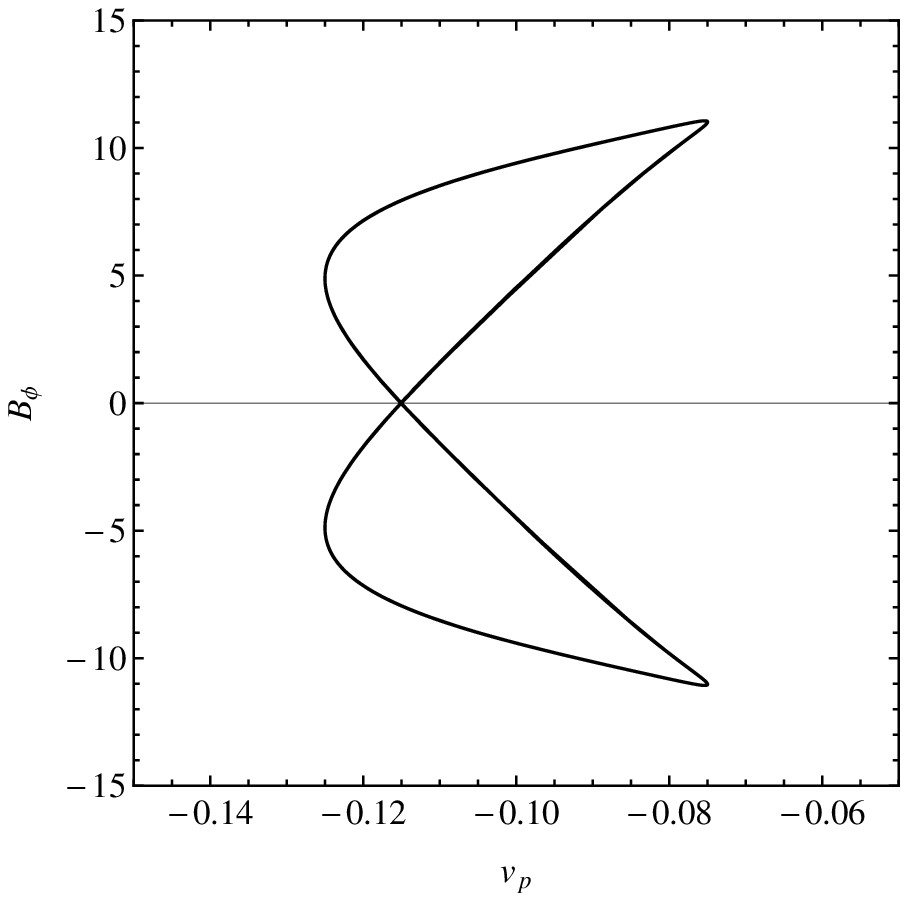}
    \includegraphics[width=40mm]{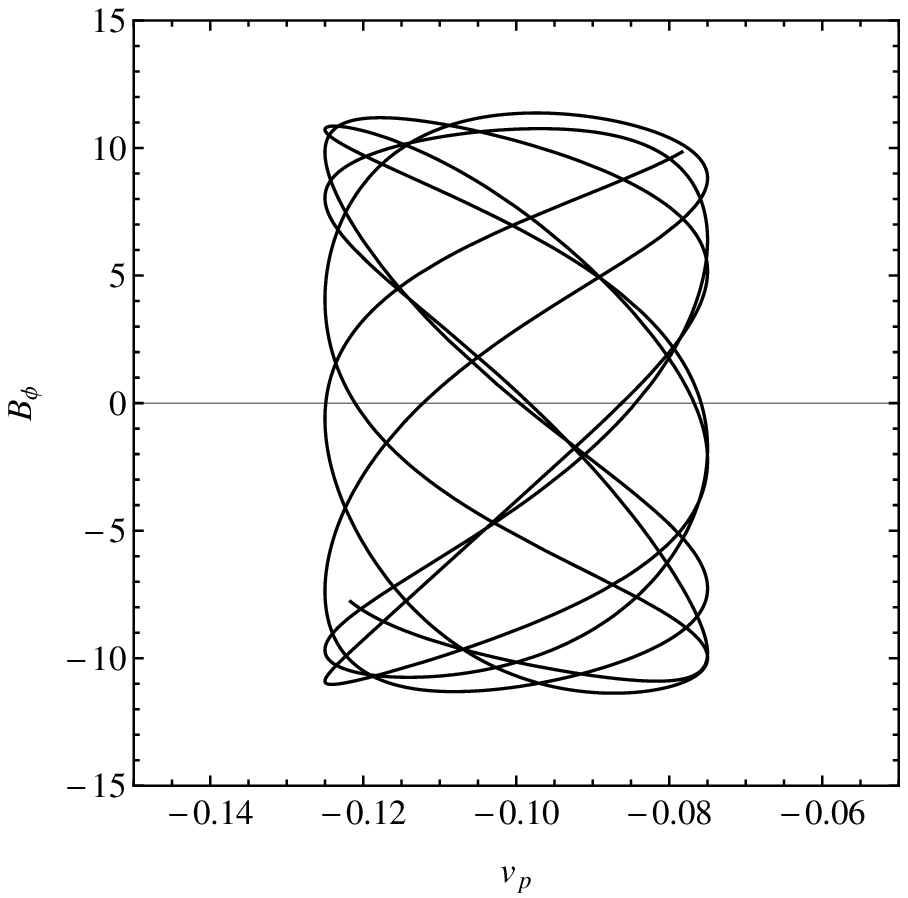}
    \caption{ \{$v_p$, $B_\phi$\} phase space for the solution between $t=1000$ and $t =1100$ using $\omega=11$ (left) and $\omega=12$ (right). The limit cycle on the left panel is observed because $\omega$ and the frequency of oscillation of $B_\phi$ is the same. }
    \label{fig-3AB}
\end{figure}

We notice that the initial phase, $\theta$, has no impact on the solution's shape (for the range of tested parameters). One interesting result is the fact that for $\omega = \frac{2\pi}{11}$ the phase difference between $v_p$ and $B_\phi^2$ is the approximately the same as the observed one, i.e., maximum amplitude of $v_p$ at the decreasing phase of the solar cycle and minimum amplitude of $v_p$ near the cycle maximum. We used as initial conditions for solving equation (\ref{eq-10}) that $B_\phi$(0)=0.01 and $dB_\phi/dt$(0)=0. For these specific initial conditions, $B_\phi(t)$ enters a steady oscillation regime approximately after t=100. For an oscillation amplitude of $v_p$ of 25\% and an initial phase $\theta=0$, the phase difference between $B_\phi(t)$ and $v_p$ "locks" around t=300 while for $\theta = \pi/2$ this occurs at approx. t=750. For different initial conditions we get different times for the "phase lock". For higher values of $v_p$ (either $v_{p0}$ or A) the time to achieve the phase lock decreases.
The phase lock occurs when the frequency $\omega$ associated with the $v_p$ fluctuation is the same as (or very close to) the natural frequency of $B_\phi$. In this low order model, the period of the cycle is given primarily by $c_2\alpha$ with a small influence of $c_1-v_p/\ell_0$ (for details c.f. \citet{Passos2012}). This small dependence on $v_p$ seems enough to ensure that after some time $B_\phi$ is synchronized with $v_p(t)$. Another way of explaining this is to resort to a \{$v_p$, $B_\phi$\} phase space. If the solution in this phase space is a limit cycle then the two quantities will synchronize phases after the solution evolves towards the attractor. See figure (\ref{fig-3AB}) for an illustrative example.

\begin{figure}%[htb!]
    \centering
    \includegraphics[width=85mm]{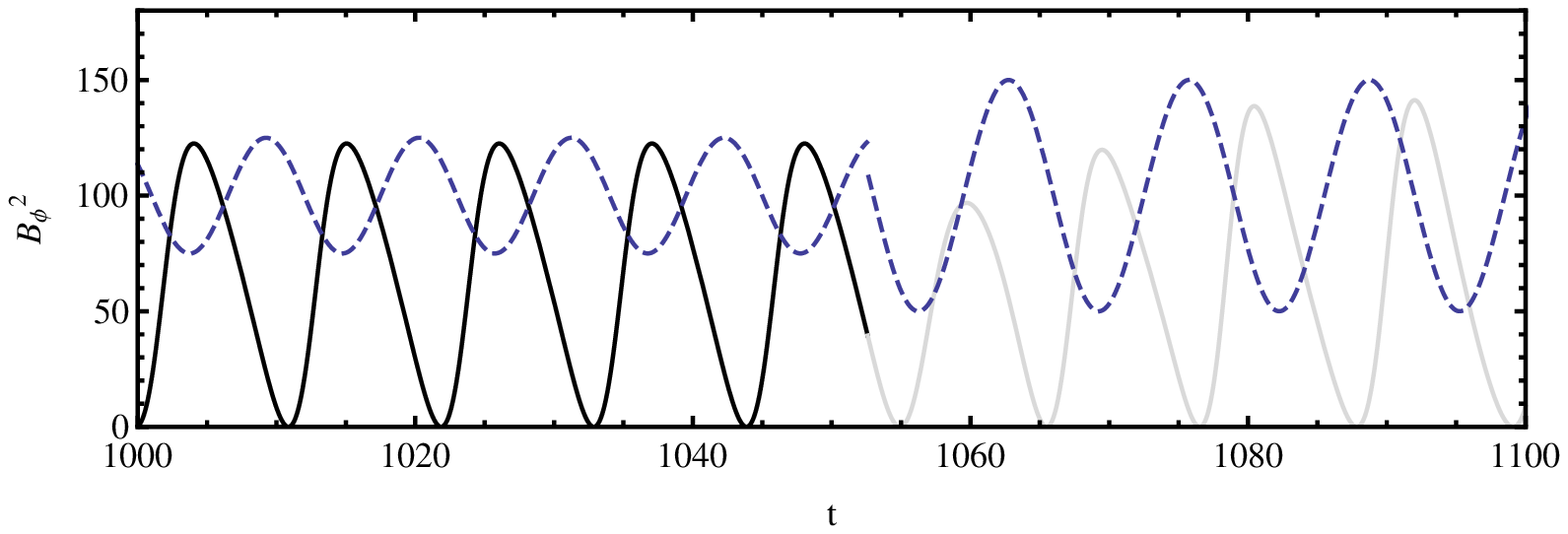}\\
    \includegraphics[width=78mm]{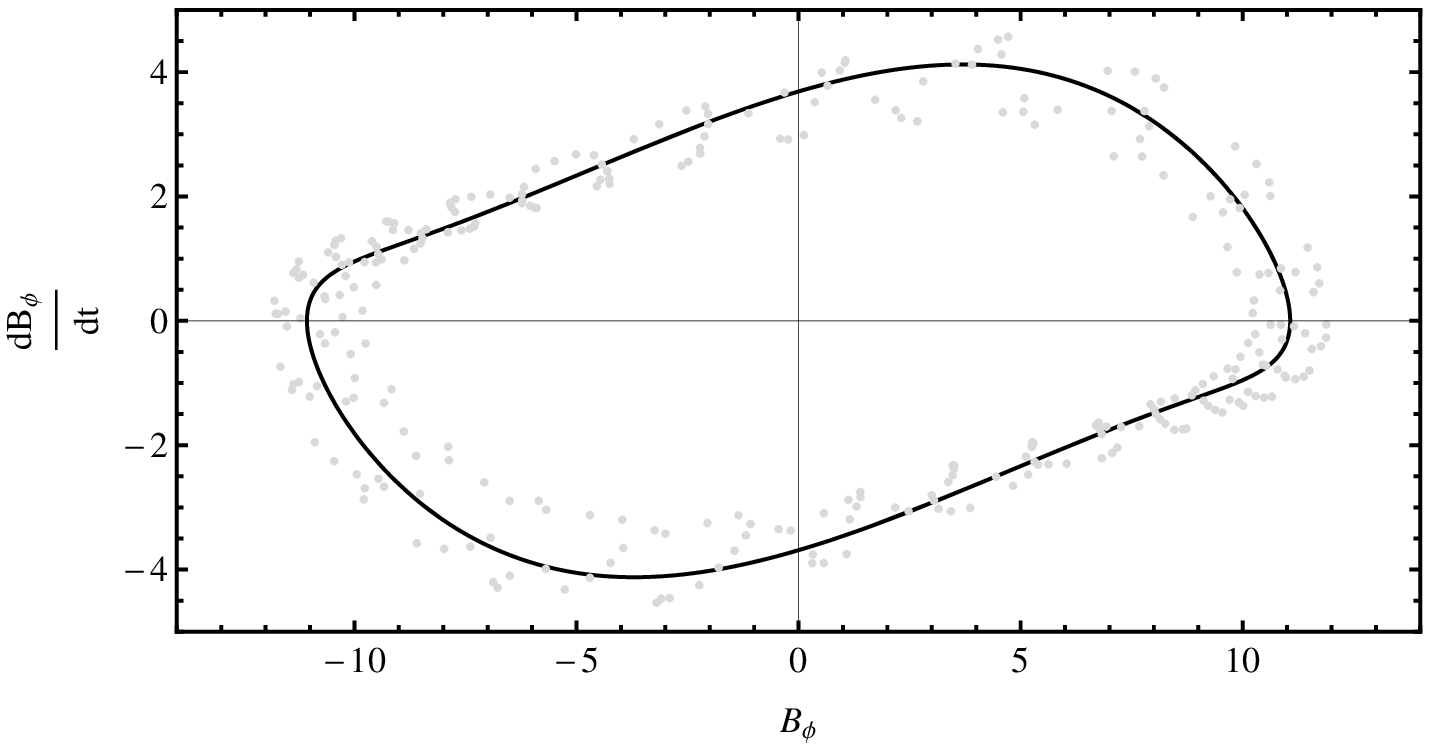}\\
    \caption{Solution assuming a variation in the fluctuation's amplitude A, and frequency, $\omega$ at a certain moment. Fixed parameters $c_1=-0.01$, $c_2\alpha=-0.095$, $c_3=0.002$, $\theta=0$, $v_{p0}$=-0.1. In the top panel we used $\omega=2\pi/11$ and A=0.025 (black) until $t =1052$ and $\omega=2\pi/13$ and A=0.05 (gray) afterwards.
    The correspondent phase space is presented in the bottom panel. The line in black corresponds to the solution between $t=1000$ and $t =1052$ and the gray dots are a sampling of the solution at regular intervals after the change.}
    \label{fig-4AB}
\end{figure}

With this set of parameters the system is well behaved and has a stable solution in the form of an attractor or limit cycle, best viewed in the phase space of $B(t)$ (figure (\ref{fig-2}) bottom panel). On the other hand different values of the fluctuation's frequency, $\omega$, yields some impact in the solution. For fluctuations with frequencies different than that of the solar cycle, like the one presented in the middle panel of figure (\ref{fig-2}), we observe a clear modulation of the field strength. This is best viewed in the corresponding phase space where we change from a stable well defined limit cycle to a "limit region". A natural variability appears in the system even if the average meridional circulation amplitude remains constant (here $v_{p0}$).
Solutions computed with different values of the fluctuation's amplitude, show that it has an influence mainly in the shape of the cycle, creating higher asymmetries between its rising and falling parts. We also observe very small changes in the frequency of the cycle but there are no signs of any long term variability (amplitude variations in the cycle's strength). With fluctuation's as high as 200\% $v_{p0}$, the solution in the phase space remains a limit cycle. According to observations, the most realistic scenario is to consider small variations in the fluctuation's amplitude and perhaps in the period (of the order of a couple of years). These variations will create a variability in the strength difference between cycles N and following cycles. In this case the observed solar cycle variability becomes dependent of changes in the fluctuations pattern of $v_p$.
We present in figure (\ref{fig-4AB}) a test case where the amplitude of the fluctuations rise from 25\% $v_{p0}$ to 50\% $v_{p0}$, and the frequency decreases from $\omega=2\pi/11$ to $\omega=2\pi/13$ at a t=1055.
In this case we find that both amplitude and frequency of the fluctuations influence the system's response. Changing the oscillation amplitude only (maintaining the frequency) results in a small variation of the limit cycle, i.e., a very smooth variability in the amplitude of the cycle appears ($\sim$0.2\%) and phase lock between flow and field is maintained. If one allows for variations in the oscillation frequency as well, a larger variability of the order of 30\% or larger appears (transition from limit cycle to limit region in the phase space). When the fluctuation's frequency occur out of phase with the solar cycle, the amplitude of the fluctuations influences the strength of the future cycles in a more pronounced way.

\section{Conclusions and final remarks}

The main objective of this work is to quickly probe the impact that cyclic fluctuations in the solar meridional circulation profile have in the dynamo process operating in our Sun. To to so, we use a simplified dynamo model where meridional circulation amplitude is forced in a sinusoidal way that mimic recent observations. We found that regular cyclic fluctuations in the amplitude of the solar meridional circulation do not seem to have an impact in the overall inner works of the solar dynamo. This is specially true when the period of these fluctuations is the same as the solar cycle's. Nevertheless, if the frequency at which the meridional flow varies becomes different from the natural frequency of the solar cycle, then the cycle becomes naturally variable. In terms of phase space of the magnetic field we go from a well behaved solution, a limit cycle attractor, to a strange attractor (or attracting region). This might be a characteristic of this specific model because here the meridional flow term, $v_p$ acts as a source term. When this source term is not synchronized with the natural frequency of the solar cycle set by $c_2\alpha$, the variability appears. In this reduced model, the relationship between cycle period and amplitude that results from changes in the meridional flow presents a non typical solar behavior. Observations show that, in average, the stronger solar cycles have shorter periods (amplitude-period rule) while in this model we get stronger cycles having longer periods. This apparent shortcoming can give us some clues about other quantities that can be varying over time. More specifically, variations in the physical mechanisms that are present in $c_2\alpha$, could produce the desired effect. Variations in the $\alpha$ effect or in the solar rotation could occur in parallel with meridional flow changes to produce the cycle's amplitude-period rule.

According to this model, amplitude variations in the fluctuation profile of $v_p$ have an impact in the shape of the solar cycle, increasing the asymmetries between rising and falling parts and even creating double peaked cycles. In this case a solar-like feature is observed, i.e., stronger cycles tend to have a steeper rising phase than weaker cycles. Large and long term variations in the amplitude of the solar cycle only occur when the frequency and amplitude of the fluctuations change in parallel.

Interesting to note is the fact that when our dynamo equation is forced with an sinusoidal meridional flow, the phase difference between both flow and field is nearly the same as the observational one. This is true as long as $c_2\alpha$ is set to reproduce the 11 year periodicity of the solar cycle. We find that for different $c_2\alpha$ values (e.g. variations in the solar rotation or in the case of other stars rotating faster or slower than the Sun) the phase difference between meridional flow and magnetic field changes. As long as both frequencies ($\omega$ and the solar cycle's) are the same (or very close) the phase difference between flow and field seems to lock independently of the initial phase $\theta$ used. Another parameter that influences the phase difference between field and flow is $c_1$. By changing the diffusivity of the system the phase lock between field and flow can be modified.

Worth mention is the fact that the exact moment, in respect to cycle N, at which the variation in the fluctuations regime occurs has an impact in the amplitude of the following cycles. Depending on the chosen variation scenario (amplitude, period or both of $v_p$) cycle N+1 can be stronger or weaker than cycle N. A similar effect as also been recently reported in 2D dynamo simulations by \cite{Nandy2011}.
We defer the details of this and other effects for a future work where we plan to perform a complete parameter space study of the model (including variations in $c_1$ and $c_3$).

From a physical point of view the information that can be extracted from such a truncated model is limited. Nevertheless we get important clues about the system's overall behavior when forced under certain parameterizations. One of the questions that this model in this present kinematic form does not address is what could be the cause(s) of the observed variations in the meridional flow. A possible explanation could be that the meridional circulation, being a weak flow, can be influenced by the Lorentz feedback from the
magnetic field. This feedback can be enough to modify significatively the meridional circulation. This scenario is supported by a recent analysis of the large-eddy global MHD simulations of the solar convection zone produced by \citet{Ghizaru2010} where \citet{PassosCharBeau2012} find evidence that the toroidal field at the base of the convection zone modulates the amplitude of the meridional circulation. If meridional circulation fluctuations are produced in this way, they should occur at the same frequency as the solar cycle. Although different variations in amplitude from cycle to cycle could occur, the phase difference between this flow and the magnetic field should remain the same. Future observation will gives us the answers.

On the other hand, if indeed the Lorentz-feedback of the field into the flow is the reason (or partially responsible) for the observed amplitude fluctuations then it is reasonable to assume that this same feedback will also influence the solar rotation, although on a smaller scale. According to the found results, variations in these two large scale flows would be more then enough to produce a variable solar cycle. More details about this physical scenario could be presented here but those would be more speculative. Our intention here is not to create/feed speculations but to explore plausible physical scenarios and motivate future studies.

The final remark that we would like present is the fact that, since numerical dynamo models are now starting to emerge as forecasting tools for solar activity, meridional variation mechanisms should be studied/implemented in order to improve their reliability. Most probably this will require a departure from the classical kinematic approach. In terms of the future, we also believe that if we keep monitoring $v_p(t)$ (and probably $\Omega(t)$) in the Sun then, according to the presented model, we would be able to predict the behavior of future solar cycles since the variability associated with the large scale flows seems to be mostly deterministic.

The authors would like to acknowledge an anonymous referee for the useful comments and the financial support of Funda\c{c}\~{a}o para a Ci\^{e}ncia e Tecnologia grant SFRH/BPD/68409/2010 and CENTRA-IST.

%%%%%%%%%%%%%%%% -- REFERENCES --  %%%%%%%%%%%%%%%%%%%%%%%%%%%%%%%%%%%%%%%

\end{document}